\newcommand{\cl}{\centerline}
\documentstyle[12pt]{article}
\begin{document}
\hfill{CCUTH-95-05}\par
\setlength{\textwidth}{6.0in}
\setlength{\textheight}{8.0in}
\setlength{\parskip}{0.0in}
\vfill
\cl{\large{{\bf Perturbative QCD Study of $B\to D^{(*)}$ Decays}}}\par
\vskip 1.0cm
\cl{Chung-Yi Wu$^1$, Tsung-Wen Yeh$^1$ and Hsiang-nan Li$^2$ }
\vskip 0.3cm
\cl{$^1$Department of Physics, National Cheng-Kung University,}
\cl{Tainan, Taiwan, R.O.C.}
\vskip 0.3cm
\cl{$^2$Department of Physics, National Chung-Cheng University,}
\cl{Chia-Yi, Taiwan, R.O.C.}
\vskip 1.0cm
\cl{\today}
\vfill
\cl{\bf Abstract}
We compute various form factors involved in $B\to D^{(*)}$ transitions
based on the perturbative QCD formalism, which includes Sudakov effects
from the resummation of large radiative corrections in a heavy-light
system. A two-parameter model wave function for $D^{(*)}$ mesons is fixed
using data of the nonleptonic decays $B\to D^{(*)}\pi$, from which the
ratio of the decay constants $f_{D^*}/f_D=0.92$ is obtained. We then derive
the spectrum of the semileptonic decay $B\to D^{*}\ell\nu$ in the fast
recoil region of the $D^*$ meson, and extract the CKM matrix element
$|V_{cb}|=0.043\times(0.12\;{\rm GeV}/f_B)\times(0.14\;{\rm GeV}/f_D)$,
$f_B$ and $f_D$ being the $B$ and $D$ meson decay constants, respectively.
Here we adopt the convention with the pion decay constant $f_\pi=93$ MeV.
With these outcomes, we evaluate the decay rate of $B\to DD_s$, and
estimate the ratio $f_{D_s}/f_D=0.98$ from data. Contributions of internal
$W$-emission and $W$-exchange diagrams are briefly discussed.

\newpage
\section{Introduction}
\vskip 0.5cm
Recently, the perturbative QCD (PQCD) formalism including Sudakov effects
has been shown to be applicable to heavy meson decays, which were usually
regarded as being dominated by nonperturbative dynamics. The breakthrough is
attributed to the resummation of large radiative corrections in heavy-light
systems such as a $B$ meson containing a light valence quark. This
resummation, which was first performed in \cite{LY1} for the semileptonic
decays $B\to \pi(\rho) \ell\nu$, improves the applicability of PQCD to
these heavy-to-light transitions. It was found that PQCD predictions are
reliable for the energy fraction of the pion above 0.3, from which the
Cabibbo-Kobayashi-Maskawa (CKM) matrix element $|V_{ub}|$ can be extracted,
once experimental data are available. A similar formalism was applied to
$B\to D$ decays in the fast recoil region of the $D$ meson \cite{L1}, and
found to be self-consistent for velocity transfer above 1.3. The resummation
technique was further employed to study the inclusive decays $B\to X_s\gamma$
and $B\to X_u\ell\nu$ \cite{L2}, and the Sudakov effects at the end points
of spectra were examined \cite{LY2}.

In this paper we shall extend the analysis in \cite{L1}, and evaluate all
the form factors involved in $B\to D^{(*)}$ transitions at the high end
of velocity transfer $\eta$. The $B$ meson wave function is determined
by the relativistic constituent quark model \cite{S}. For fast $D^{(*)}$
mesons, a convincing model wave function is still not yet obtained. We
propose a two-parameter $D^{(*)}$ meson wave function, and fix the parameters
using data in the large $\eta$ region from
the nonleptonic decays $B\to D^{(*)}\pi$ \cite{A}. One of the
parameters corresponds to the normalization constant, and the other
controls the shape. With these phenomenological inputs, our analysis
is free of the ambiguity from nonperturbative effects. We compare our
results to those from other
theoretical approaches, for example, heavy quark symmetry (HQS) \cite{IW}
combined with ${\cal O}(\alpha_s)$ and ${\cal O}(1/M)$ corrections in
\cite{N} and overlap integrals of heavy meson wave functions
in \cite{NR,AGS}.

We derive the spectrum of the semileptonic decay $B\to D^*\ell\nu$ from
the transition form factors. The CKM matrix element $|V_{cb}|$ is then
obtained by
fitting our predictions to data \cite{B} at large $\eta$, where PQCD is
reliable. Our approach to the extraction of $|V_{cb}|$ differs from those
in the literature, where the behaviors of the form factors at $\eta=1$ are
employed. The relevant nonvanishing form factors take the same functional
form $\xi(\eta)$, which is normalized to unity at zero recoil,
$\xi(\eta=1)=1$, because of HQS. The quantity $|V_{cb}|\xi(1)$ can be
extracted from experimental data, if the behavior of $\xi$ above zero recoil
is known. However, $\xi$ is thought of as being uncalculable in perturbation
theory, and thus the extraction depends on how to extrapolate $\xi$ from
$\eta=1$ to $\eta>1$. Hence, different models of $\xi(\eta)$ lead to
different values of $|V_{cb}|$ \cite{N,B}. We argue that $\xi$ is in fact
calculable in the fast recoil region. The nonperturbative wave functions,
fixed by data of other decay modes, provide model-independent extraction of
$|V_{cb}|$.

The consistency of $|V_{cb}|$ determined in this paper with currently
accepted values justifies the PQCD analysis of $B$ meson decays, especially
$B\to\pi$ decays. Predictions for the spectra of the decays $B\to\pi(\rho)
\ell\nu$ in \cite{LY1} are then convincing, and can be used to extract
$|V_{ub}|$. On the other hand, HQS requires only the normalization of
heavy-to-heavy transition form factors at zero recoil. The PQCD formalism,
however, gives information near the high end of $\eta$. Therefore, these two
approaches complement each other. Furthermore, our formalism is applicable
to the evaluation of nonfactorizable contributions from $W$-exchange
diagrams, which remains a challenging subject in the study of nonleptonic
$B$ meson decays.

We develop the factorization formulas for all the $B\to D^{(*)}$ transition
form factors in Sect. 2. In Sect. 3 the two-parameter $D^{(*)}$ meson wave
function is determined by fitting the data of the decays $B\to D^{(*)}\pi$.
The $B\to D^*\ell\nu$ spectrum is derived, and $|V_{cb}|$ is extracted from
experimental data \cite{B}. We show in Sect. 4 that nonfactorizable
contributions from $W$-exchange diagrams are negligible. Sect. 5 is the
conclusion.
\vskip 2.0cm

\section{Factorization Formulas}
\vskip 0.5cm

The six form factors $\xi_i$, $i=+$, $-$, $V$, $A_1$, $A_2$ and $A_3$,
involved in $B\to D^{(*)}$ transitions are defined by the following matrix
elements:
\begin{eqnarray}
\langle D (P_2)|V^\mu|B(P_1)\rangle
&=&\sqrt{M_BM_D}(\xi_+(\eta)(v_1+v_2)^\mu+
\xi_-(\eta)(v_1-v_2)^\mu)\;,
\nonumber \\
\langle D^* (P_2)|V^\mu|B(P_1)\rangle
&=&i\sqrt{M_BM_{D^*}}\xi_V(\eta)
\epsilon^{\mu\nu\alpha\beta}\epsilon^*_\nu v_{2\alpha}v_{1\beta}\;,
\nonumber \\
\langle D^* (P_2)|A^\mu|B(P_1)\rangle&=&\sqrt{M_BM_{D^*}}
[\xi_{A_1}(\eta)(\eta+1)\epsilon^{*\mu}
-\xi_{A_2}(\eta)\epsilon^*\cdot v_1 v_1^\mu
\nonumber \\
& &-\xi_{A_3}(\eta)\epsilon^*\cdot v_1 v_2^\mu]\;.
\label{iwm}
\end{eqnarray}
The momentum $P_1$ $(P_2)$, the mass $M_B$ ($M_{D^{(*)}}$) and the velocity
$v_1$ $(v_2)$ of the $B$ $(D^{(*)})$ meson are related by $P_1=M_Bv_1$
$(P_2=M_{D^{(*)}}v_2)$. The velocity transfer $\eta=v_1\cdot v_2$ has been
introduced before, whose expression in terms of the momentum transfer
$q^2=(P_1-P_2)^2$ is given by
\begin{equation}
\eta=\frac{M_B^2+M_{D^{(*)}}^2-q^2}{2M_BM_{D^{(*)}}}\;.
\end{equation}
$\epsilon^*$ is the polarization vector of the $D^*$ meson, satisfying
$\epsilon^*\cdot v_2=0$. The vector current $V^\mu$ and the axial vector
current $A^\mu$ are defined by $V^\mu={\bar c}\gamma^{\mu}b$ and
$A^\mu={\bar c}\gamma^{\mu}\gamma_5 b$, respectively.

In the infinite mass limit of $M_B$ and $M_{D^{(*)}}$, the six form factors
have the relations
\begin{equation}
\xi_+=\xi_V=\xi_{A_1}=\xi_{A_3}=\xi,\;\;\;\;  \xi_-=\xi_{A_2}=0,
\label{iwr}
\end{equation}
where $\xi$ is the Isgur-Wise (IW) function \cite{IW} mentioned in the
Introduction. $\xi$ is normalized to unity at zero recoil from HQS.
For the behavior of $\xi$ above zero recoil, there is only model
estimation from the overlap integrals of heavy meson wave functions
\cite{NR,AGS}.

We work in the rest frame of the $B$ meson, in which $P_1$ is written,
using light-cone components, as $P_1=M_B/\sqrt{2}(1,1,{\bf 0}_T)$, and
$P_2$ has the nonvanishing components \cite{L1}
\begin{eqnarray}
P_2^+=\frac{\eta+\sqrt{\eta^2-1}}{\sqrt{2}}M_{D^{(*)}}\;,
\nonumber \\
P_2^-=\frac{\eta-\sqrt{\eta^2-1}}{\sqrt{2}}M_{D^{(*)}}\;.
\end{eqnarray}
As $\eta\to 1$ with $P_2^+=P_2^-=M_{D^{(*)}}/\sqrt{2}$,
the $D^{(*)}$ meson behaves like a heavy meson. However, in the large $\eta$
limit with $P_2^+\gg M_{D^{(*)}}/\sqrt{2}\gg P_2^-$, the $D^{(*)}$ meson can
be regarded as being light \cite{L1}. We argue that
$\xi$ is dominated by soft contributions in the slow $D^{(*)}$ meson limit,
where the heavy meson wave functions strongly overlap, and factorization
theorems fail. However, when the $D^{(*)}$ meson recoils fast, carrying
energy much greater than $M_{D^{(*)}}$, $B\to D^{(*)}$ transitions are then
similar to $B\to\pi$ ones \cite{L1} as stated above, and PQCD is expected to
be applicable. In this paper we shall show that the PQCD formalism including
Sudakov effects gives reliable predictions for $\xi$ in the large $\eta$
region.

The PQCD factorization formulas for the $B\to D$ transition form
factors have been derived in \cite{L1}. Here we summarize the idea.
The factorization of the matrix elements in eq.~(\ref{iwm}) into the
convolution of a hard scattering amplitude with the $B$ and $D^{(*)}$ meson
wave functions is shown in fig.~1a, where the $b$ and $c$ quarks are
represented by the thicker and thick lines, respectively. $k_1$ ($k_2$) is
the momentum of the light valence quark in the $B$ ($D^{(*)}$) meson,
satisfying $k_1^2\approx 0$ ($k_2^2\approx 0$). $k_1$ has a large minus
component $k_1^-$, which defines the momentum fraction $x_1=k_1^-/P_1^-$,
and small transverse components ${\bf k}_{1T}$, which serve as the infrared
cutoff of loop integrals for radiative corrections. Similarly, $k_2$ has
a large component $k_2^+$, defining $x_2=k_2^+/P_2^+$, and small
${\bf k}_{2T}$.

A simple investigation shows that large logarithms arise from radiative
corrections to the above factorization picture. In particular, double
(leading) logarithms occur in the reducible corrections illustrated by
the ${\cal O}(\alpha_s)$ diagrams in fig.~1b \cite{LY1}, when they are
evaluated in axial gauge. In order to have a reliable PQCD analysis, these
double logarithms must be organized using the resummation technique, which
leads to the evolution of the $B$ ($D^{(*)}$) meson wave function $\phi_B$
($\phi_{D^{(*)}}$) in $k_1^-$ ($k_2^+$). We quote the results as follows
\cite{L1}:
\begin{eqnarray}
\phi_B(x_1,P_1,b_1,\mu)&\approx&\phi_B(x_1)\exp\left[-s(k_1^-,b_1)-2
\int_{1/b_1}^{\mu}\frac{d{\bar \mu}}{\bar \mu}\gamma(\alpha_s({\bar \mu}))
\right]\;,
\nonumber \\
\phi_{D^{(*)}}(x_2,P_2,b_2,\mu)&\approx&\phi_{D^{(*)}}(x_2)\exp
\biggl[-s(k_2^+,b_2)-s(P_2^+-k_2^+,b_2)
\nonumber \\
& &\left.-2\int_{1/b_2}^{\mu}
\frac{d{\bar \mu}}{\bar \mu}\gamma(\alpha_s({\bar \mu}))\right]\;,
\label{wp}
\end{eqnarray}
where $b_1$ ($b_2$) is the Fourier conjugate variable to $k_{1T}$
($k_{2T}$), and can be regarded as the spatial extent of the $B$ ($D^{(*)}$)
meson. $\mu$ is the renormalization and factorization scale. The exponent
$s$ organizes the double logarithms from the overlap of collinear and soft
divergences, and the integral, with the quark anomalous dimension
$\gamma=-\alpha_s/\pi$ as the integrand, groups the remaining single
ultraviolet logarithms in fig.~1b. The expression of $s$ is very
complicated, and exhibited in the Appendix.

In the resummation procedures the $B$ meson is treated as a heavy-light
system. The $D^{(*)}$ meson is, however, treated as a light-light system as
indicated in eq.~(\ref{wp}), because we concentrate on the fast recoil
region \cite{L1}. The initial conditions $\phi_i(x)$ of the evolution,
$i=B$, $D$ and $D^*$, are of nonperturbative origin, satisfying the
normalization
\begin{equation}
\int_0^1\phi_i(x)dx=\frac{f_i}{2\sqrt{3}}\;,
\label{no}
\end{equation}
with $f_i$ the meson decay constants. The initial condition in eq.~(\ref{wp})
should be written as $\phi(x,b,1/b)$, and $\phi(x)$ is in fact
an approximation. We have neglected the
intrinsic transverse momentum dependence denoted by the argument $b$, and
PQCD corrections proportional to $\alpha_s(1/b)$, because these two effects
cancel partially \cite{LY1}.

The evolution of the hard scattering amplitude $H$ from the summation of
single ultraviolet logarithms is expressed as \cite{L1}
\begin{equation}
H(k_1^-,k_2^+,b_1, b_2,\mu)\approx H^{(0)}(k_1^-,k_2^+,b_1,b_2,t)\exp
\left[-4\int_{\mu}^t\frac{d{\bar \mu}}{\bar \mu}\gamma(\alpha_s({\bar
\mu}))\right]\;,
\label{eh}
\end{equation}
where the variable $t$ denotes the largest mass scale of $H$. We have
approximated $H$ by the ${\cal O}(\alpha_s)$ expression $H^{(0)}$, which
makes sense if perturbative contributions indeed dominate.
Combining the evolution in the above formula, we obtain the complete Sudakov
factor $e^{-S}$, where the exponent $S$ is given by
\begin{eqnarray}
S(k_1^-,k_2^+,b_1,b_2)&=&s(k_1^-,b_1)+s(k_2^+,b_2)+s(P_2^+-k_2^+,b_2)
\nonumber \\
& &-\frac{1}{\beta_1}\left[\ln\frac{\ln(t/\Lambda)}{-\ln(b_1\Lambda)}
+\ln\frac{\ln(t/\Lambda)}{-\ln(b_2\Lambda)}\right]
\label{S}
\end{eqnarray}
with $\beta_1=(33-2n_f)/12$, $n_f=4$ being the flavor number. The QCD
scale $\Lambda\equiv\Lambda_{\rm QCD}$ will be set to 0.2 GeV below. It is
easy to show that $e^{-S}$ falls off quickly in the large $b$, or
long-distance, region, giving so-called Sudakov suppression.

With the above brief discussion, the only thing left is to compute $H^{(0)}$
for each form factor. The calculation of $H^{(0)}$ for $\xi_\pm$ has been
performed in \cite{L1}. In a similar way we derive $H^{(0)}$ for other
$\xi$'s. The factorization formulas for all the transition form factors in
$b$ space, with Sudakov suppression included, are listed below:
\begin{eqnarray}
\xi_+&=& 16\pi{\cal C}_F\sqrt{M_BM_D}
\int_{0}^{1}d x_{1}d x_{2}\,\int_{0}^{\infty} b_1d b_1 b_2d b_2\,
\phi_B(x_1)\phi_D(x_2)
\nonumber \\
& &\times [(M_B+x_2\zeta_+M_D)h(x_1,x_2,b_1,b_2)
+(M_D+x_1\zeta_+M_B)h(x_2,x_1,b_2,b_1)]
\nonumber \\
& &\times \exp[-S(k_1^-,k_2^+,b_1,b_2)]\;,
\label{+}
\\
\xi_-&=& -16\pi{\cal C}_F\sqrt{M_BM_D}
\int_{0}^{1}d x_{1}d x_{2}\,\int_{0}^{\infty} b_1d b_1 b_2d b_2\,
\phi_B(x_1)\phi_D(x_2)
\nonumber \\
& &\times \zeta_-[x_2M_Dh(x_1,x_2,b_1,b_2)
-x_1M_Bh(x_2,x_1,b_2,b_1)]
\nonumber \\
& &\times \exp[-S(k_1^-,k_2^+,b_1,b_2)]\;,
\label{-}
\\
\xi_V&=& 16\pi{\cal C}_F\sqrt{M_BM_{D^*}}
\int_{0}^{1}d x_{1}d x_{2}\,\int_{0}^{\infty} b_1d b_1 b_2d b_2\,
\phi_B(x_1)\phi_{D^*}(x_2)
\nonumber \\
& &\times [(M_B-x_2\zeta_1M_{D^*})h(x_1,x_2,b_1,b_2)
+(M_{D^*}+x_1\zeta_2M_B)h(x_2,x_1,b_2,b_1)]
\nonumber \\
& &\times \exp[-S(k_1^-,k_2^+,b_1,b_2)]\;,
\label{V}
\\
\xi_{A_1}&=& 16\pi{\cal C}_F\sqrt{M_BM_{D^*}}
\int_{0}^{1}d x_{1}d x_{2}\,\int_{0}^{\infty} b_1d b_1 b_2d b_2\,
\phi_B(x_1)\phi_{D^*}(x_2)
\nonumber \\
& &\times [(M_B-x_2\zeta_3M_{D^*})h(x_1,x_2,b_1,b_2)
+(M_{D^*}+x_1\zeta_4M_B)h(x_2,x_1,b_2,b_1)]
\nonumber \\
& &\times \exp[-S(k_1^-,k_2^+,b_1,b_2)]\;,
\label{A1}
\\
\xi_{A_2}&=& -16\pi{\cal C}_F\sqrt{M_BM_{D^*}}
\int_{0}^{1}d x_{1}d x_{2}\,\int_{0}^{\infty} b_1d b_1 b_2d b_2\,
\phi_B(x_1)\phi_{D^*}(x_2)
\nonumber \\
& &\times x_1\zeta_5M_B h(x_2,x_1,b_2,b_1)
\exp[-S(k_1^-,k_2^+,b_1,b_2)]\;,
\label{A2}
\\
\xi_{A_3}&=& \xi_V\;,
\label{A3}
\end{eqnarray}
with the constants
\begin{eqnarray}
& &\zeta_+=\frac{1}{2}\left[\eta-\frac{3}{2}+
\sqrt{\frac{\eta-1}{\eta+1}}\left(\eta-\frac{1}{2}\right)\right]\;,
\nonumber \\
& &\zeta_-=\frac{1}{2}\left[\eta-\frac{1}{2}+
\sqrt{\frac{\eta+1}{\eta-1}}\left(\eta-\frac{3}{2}\right)\right]\;,
\nonumber \\
& &\zeta_1=\frac{1}{2}+\frac{\eta-2}{2\sqrt{\eta^2-1}}\;,
\nonumber \\
& &\zeta_2=\frac{1}{2\sqrt{\eta^2-1}}\;,
\nonumber \\
& &\zeta_3=\frac{2-\eta-\sqrt{\eta^2-1}}{\eta+1}\;,
\nonumber \\
& &\zeta_4=\frac{1}{2(\eta+1)}\;,
\nonumber \\
& &\zeta_5=1+\frac{\eta}{\sqrt{\eta^2-1}}\;.
\end{eqnarray}
${\cal C}_F=4/3$ is the color factor. The function $h$, coming from the
Fourier transform of $H^{(0)}$, is given by
\begin{eqnarray}
h(x_1,x_2,b_1,b_2)&=&
\alpha_{s}(t)K_{0}\left(\sqrt{x_1x_2\zeta M_BM_{D^{(*)}}}b_1\right)
\nonumber \\
& &\times \left[\theta(b_1-b_2)K_0\left(\sqrt{x_2\zeta M_BM_{D^{(*)}}}
b_1\right)I_0\left(\sqrt{x_2\zeta M_BM_{D^{(*)}}}b_2\right)\right.
\nonumber \\
& &\;\;\;\;\left.
+\theta(b_2-b_1)K_0\left(\sqrt{x_2\zeta M_BM_{D^{(*)}}}b_2\right)
I_0\left(\sqrt{x_2\zeta M_BM_{D^{(*)}}}b_1\right)\right]
\nonumber\\
& &
\label{dh}
\end{eqnarray}
with the constant $\zeta=\eta+\sqrt{\eta^2-1}$. The scale $t$ is chosen as
\cite{L1}
\begin{equation}
t={\rm max}(\sqrt{x_1x_2\zeta M_BM_{D^{(*)}}},1/b_1,1/b_2)\;.
\end{equation}
Note the equality of $\xi_V$ and $\xi_{A_3}$. We argue that $\xi_V$ will
differ from $\xi_{A_3}$, if higher-order corrections to the initial
condition $H(k_1^-,k_2^+,b_1,b_2,t)$ are taken into account.
\vskip 2.0cm

\section{Determination of $|V_{cb}|$}
\vskip 0.5cm

Before proceeding to the numerical analysis of eqs.~(\ref{+})-(\ref{A3}),
we discuss how to fix the $B$ and $D^{(*)}$ meson wave functions. For the
$B$ meson, we adopt the wave function from the relativistic constituent
quark model \cite{S}, $\phi_B(x)=\int d^2{\bf k}_T/(4\pi)^2 \phi_B
(x,{\bf k}_T)$, with
\begin{equation}
\phi_B(x,{\bf k}_T)=N_B\left[C_B+\frac{M_B^2}{1-x}+\frac{k_T^2}{x(1-x)}
\right]^{-2}\;.
\end{equation}
The normalization constant $N_B$ and the shape parameter $C_B$ are
determined by the conditions
\begin{eqnarray}
\int_0^1dx\int \frac{d^2{\bf k}_T}{(4\pi)^2} \phi_B(x,{\bf k}_T)&=&
\frac{f_B}{2\sqrt{3}}\;,
\nonumber \\
\int_0^1dx\int \frac{d^2{\bf k}_T}{(4\pi)^2} [\phi_B(x,{\bf k}_T)]^2&=&
\frac{1}{2}\;.
\label{cs}
\end{eqnarray}
We obtain $N_B=49.5$ GeV$^3$ and $C_B=-27.699845$ GeV$^2$ for
$M_B=5.28$ GeV \cite{PDG} and the $B$ meson decay constant $f_B=0.12$ GeV
from the lattice calculation \cite{BLS}. The $B$ meson wave function is
then given by
\begin{equation}
\phi_B(x)=\frac{N_B}{16\pi}\frac{x(1-x)^2}{M_B^2+C_B(1-x)}\;.
\label{bw}
\end{equation}

For the $D^{(*)}$ meson, we propose the following model, which possesses
the same functional form as eq.~(\ref{bw}),
\begin{equation}
\phi_{D^{(*)}}(x)=\frac{N_{D^{(*)}}}{16\pi}\frac{x(1-x)^2}
{M_{D^{(*)}}^2+C_{D^{(*)}}(1-x)}\;.
\label{dw}
\end{equation}
Equation~(\ref{cs}) is not appropriate for the determination of $N_{D^{(*)}}$
and $C_{D^{(*)}}$, when the $D^{(*)}$ meson recoils fast \cite{L1}. Hence,
we shall take an alternative approach. It is known from PQCD factorization
theorems that wave functions are universal, or process-independent. It hints
that we can fix the parameters using data of any $B\to D^{(*)}$ decays,
such as the two-body nonleptonic decays $B\to D^{(*)}\pi$, for which
factorization theorems should work best. With these phenomenological
inputs at specific kinematic points $\eta=\eta_{\rm max}$, we then predict
the behaviors of all $\xi$'s in a finite range of $\eta$.

We assume the vertex factorization hypothesis for the following analysis,
which has been shown to be consistent with current experimental data
\cite{BSS}. We list the formulas for the decay rates of various
$B\to D^{(*)}$ decay modes involved in our study. They are
\begin{eqnarray}
\Gamma({\bar B}^0\to D^+\pi^-)&=&\frac{1}{32\pi}G_F^2|V_{cb}|^2|V_{ud}|^2
f_\pi^2M_B^3\frac{(1-r^2)^3(1-r)^2}{2r}
\nonumber \\
& &\times\left[\frac{1+r}{1-r}\xi_+
(\eta_{\rm max})-\xi_-(\eta_{\rm max})\right]^2\;,
\nonumber \\
\Gamma({\bar B}^0\to D^{*+}\pi^-)&=&\frac{1}{32\pi}G_F^2|V_{cb}|^2|V_{ud}|^2
f_\pi^2M_B^3\frac{(1-r^2)^5}{(2r)^3}
\nonumber \\
& &\times\left[\frac{1+r}{1-r}\xi_{A_1}
(\eta_{\rm max})-(r\xi_{A_2}(\eta_{\rm max})+\xi_{A_3}(\eta_{\rm max}))
\right]^2\;,
\nonumber \\
& &
\label{td}
\end{eqnarray}
for the nonleptonic decays, with the constants $r=M_{D^{(*)}}/M_B$ and
$\eta_{\rm max}=(1+r^2)/(2r)$, and
\begin{eqnarray}
\frac{d \Gamma}{d q^2}&=&
\frac{1}{96\pi^3}G_F^2|V_{cb}|^2M_B^3r^2(\eta^2-1)^{1/2}(\eta+1)^2
\nonumber\\
& &\times\left\{2(1-2\eta r+r^2)\left[\xi_{A_1}^2(\eta)+\frac{\eta-1}{\eta+1}
\xi_V^2(\eta)\right]\right.
\nonumber \\
& &+\left[(\eta-r)\xi_{A_1}(\eta)-(\eta-1)\left(r\xi_{A_2}(\eta)+
\xi_{A_3}(\eta)\right)\right]^2\Biggr\}
\label{dd}
\end{eqnarray}
for the spectrum of the semileptonic decay
${\bar B}^0\to D^{*+}\ell^-{\bar \nu}$ \cite{N}. For nonleptonic decays,
there exist additional important corrections from final-state interactions
with soft gluons attaching the $D^{(*)}$ meson and the pion. It has been
argued that these corrections produce only single logarithms, which cancel
asymptotically \cite{L1}, and are thus not considered here.

We adopt $G_F=1.16639\times 10^{-5}$ GeV$^{-2}$ for the Fermi coupling
constant, $|V_{ud}|=0.974$ for the CKM matrix element, $M_D=1.87$ GeV and
$M_{D^*}=2.01$ GeV for the $D$ and $D^*$ meson masses \cite{PDG},
respectively, $\tau_{B^0}=1.53$ ps for the ${\bar B}^0$ meson lifetime
\cite{B}, $f_\pi=93$ MeV for the pion decay constant and $f_D=0.14$ GeV
for the $D$ meson decay constant from the lattice calculation
\cite{BLS}. The four parameters we shall determine are $C_D$ ($N_D$ is
fixed by $f_D$ from eq.~(\ref{no})), $C_{D^*}$, $f_{D^*}$ (or $N_{D^*}$
equivalently) and $|V_{cb}|$. At the same time, we have four
constraints from experimental data: the branching ratios
${\cal B}({\bar B}^0\to D^+\pi^-)=2.9\times 10^{-3}$ and
${\cal B}({\bar B}^0\to D^{*+}\pi^-)=2.6\times 10^{-3}$ \cite{A}, and the
height and the profile of $d\Gamma/dq^2$ at large $\eta$, or equivalently,
its values at $q^2=M_{\pi^+}^2\approx 0$ and at $q^2=M_{D_s}^2$ \cite{B},
$M_{D_s}=1.97$ GeV being the $D_s$ meson mass \cite{PDG}. In principle, we
can determine the $D^{(*)}$ meson wave functions and the CKM matrix element
$|V_{cb}|$ completely from the data fitting.

If there are more data from other decay modes, such as $B\to D^{(*)}D_s$,
$f_D$ can be fixed phenomenologically, and needs not to be specified at the
beginning. However, these data still suffer large errors \cite{ST}, and it
is not practical to perform the fitting based on them. On the other hand,
the data of $B\to\rho$ decays, ${\cal B}({\bar B}^0\to D^+\rho^-)=
8.1\times 10^{-3}$ and ${\cal B}({\bar B}^0\to D^{*+}\rho^-)=7.4\times
10^{-3}$ \cite{A}, do not give more information than $B\to\pi$ decays,
and are not employed here. This is obvious from the equality of the ratios
\begin{equation}
\frac{{\cal B}({\bar B}^0\to D^+\rho^-)}{{\cal B}({\bar B}^0\to
D^{*+}\rho^-)}\approx
\frac{{\cal B}({\bar B}^0\to D^+\pi^-)}{{\cal B}({\bar B}^0\to
D^{*+}\pi^-)}\approx 1.1\;.
\label{eq}
\end{equation}
Hence, the data of $B\to \rho$ decays just lead to the $\rho$ meson
decay constant $f_\rho\approx f_\pi\times \sqrt{8.1/2.9}=0.155$ GeV,
consistent with the currently accepted value. Equation~(\ref{eq}) is a
direct consequence of the vertex factorization hypothesis for the negligible
$\rho$ meson mass. Note that the Bauer-Stech-Wirbel (BSW) method \cite{BSW}
gives different predictions from eq.~(\ref{eq}), which are \cite{A}
\begin{equation}
\frac{{\cal B}({\bar B}^0\to D^+\rho^-)}{{\cal B}({\bar B}^0\to
D^{*+}\rho^-)}=0.885\;,\;\;\;\;
\frac{{\cal B}({\bar B}^0\to D^+\pi^-)}{{\cal B}({\bar B}^0\to
D^{*+}\pi^-)}= 1.04\;.
\label{eq1}
\end{equation}

Our plan is summarized as follows:

\noindent
1. Assume a set of initial values of $C_D$ and $C_{D^*}$, say,
$C_D=-2.9$ GeV$^2$ \cite{L1} and $C_{D^*}=-3.4$ GeV$^2$. Determine the value
of $f_{D^*}$ from the ratio ${\cal B}({\bar B}^0\to D^+\pi^-)/{\cal B}
({\bar B}^0\to D^{*+}\pi^-)=1.1$, which is independent of $|V_{cb}|$.

\noindent
2. Extract $|V_{cb}|$ from the magnitude of the spectrum $d\Gamma/dq^2$
at $q^2=M_{\pi^+}^2$ using $f_{D^*}$ from Step 1, and the initial $C_D$ and
$C_{D^*}$.

\noindent
3. Determine a new value of $C_D$ from ${\cal B}({\bar B}^0\to D^+\pi^-)=
2.9\times 10^{-3}$, and a new $C_{D^*}$ from $d\Gamma/dq^2$ at
$q^2=M_{D_s}^2$ using $f_{D^*}$ and $|V_{cb}|$ obtained from the above two
steps.

\noindent
4. Go to Step 1, starting with the new initial values of $C_D$ and
$C_{D^*}$.

\noindent
At last, the four parameters approach their limits after a few iterations.
The results are
\begin{eqnarray}
& &C_D=-2.6179\;\;{\rm GeV}^2\;,\;\;\;N_D=13.8\;\;{\rm GeV}^3\;,
\;\;\;f_D=0.14\;\;{\rm GeV}\;,
\nonumber \\
& &C_{D^*}=-3.0421\;\;{\rm GeV}^2\;,\;\;\; N_{D^*}=14.6\;\;
{\rm GeV}^3\;,\;\;\;f_{D^*}=0.129\;\;{\rm GeV}\;,
\nonumber \\
& &|V_{cb}|=0.043\;.
\label{fp}
\end{eqnarray}

The dependence of the $B$ and $D^{(*)}$ meson wave functions on the momentum
fraction $x$ is shown in fig.~2. $\phi_B$ peaks at $x\approx 0.05$, and
$\phi_{D^{(*)}}$ peaks at $x\approx 0.2$, indicating that the heavier
$B$ meson is strongly dominated by soft dynamics. The profiles of $\phi_D$
and $\phi_{D^*}$ are very similar, but the maximum of $\phi_{D^*}$ locates
at a slightly smaller $x$ compared to $\phi_D$ because of the relation
$|C_{D^*}|/M_{D^*}^2 > |C_D|/M_D^2$. All these features are consistent
with the expectation from the ordering of the
masses, $M_B\gg M_{D^*}>M_D$.  That the height of $\phi_D$ is larger than
that of $\phi_{D^*}$ is due to $f_D>f_{D^*}$. Note that our prediction
$f_{D^*}/f_D=0.92$ is contrary to  $f_{D^*}/f_D=1.28$ appearing
in the literature \cite{AN}.

The parameters in eq.~(\ref{fp}) lead to the branching ratios
${\cal B}({\bar B}^0\to D^+\pi^-)=2.89\times 10^{-3}$ and
${\cal B}({\bar B}^0\to D^{*+}\pi^-)=2.61\times 10^{-3}$, and the spectrum
$d\Gamma/dq^2$ for the semileptonic decay
${\bar B}^0 \to D^{*+}\ell^-{\bar \nu}$ as in fig.~3, where the
CLEO data from Ref.~\cite{B} are also shown. It is observed that our
predictions are in good agreement with the data at low $q^2$, and begin to
deviate above $q^2=4$ GeV$^2$, the slow recoil region in which PQCD is not
reliable. In order to justify the PQCD analysis for $q^2<4$ GeV$^2$, or
$\eta>1.3$ approximately, we exhibit the dependence of $\xi_+$ and $\xi_{A_1}$
on the cutoff $b_c=b_1=b_2$ for $\eta=1.3$ in fig.~4. More than 50\% of the
contributions to the form factors come from the region with $b<0.6/\Lambda$,
{\it ie.}, $\alpha_s(1/b_c)/\pi< 0.5$.

Variation of the six transition form factors with the velocity transfer
$\eta$ is displayed in fig.~5. From fig.~5a, we find that the magnitudes
of $\xi_V$, $\xi_{A_1}$ and $\xi_{A_3}$ are almost equal, with the relation
$\xi_V=\xi_{A_3}>\xi_{A_1}$, and their behaviors are close to that of
$\xi_+$, corresponding to the similarity between the profiles of
$\phi_{D^*}$ and $\phi_D$. This similarity is also reflected by the fact
that the ratio of $\xi_V$ to $\xi_+$ is roughly the same as $f_{D^*}/f_D$.
Contrary to fig.~5a, $\xi_-$ and $\xi_{A_2}$ shown in
fig.~5b increase with $\eta$. $\xi_-$ possesses a smaller slope, and is
expected to become negative at low $\eta$. These features are consistent
with the predictions from HQS combined with ${\cal O}
(\alpha_s)$ and ${\cal O}(1/M)$ corrections in \cite{N}.

The CKM matrix element $|V_{cb}|=0.043$ extracted here
is a bit larger than recent estimations in the literature, which range
from 0.035 to 0.040 \cite{AGS,B,GGL}. Refer to Ref.~\cite{B} in which
$\xi_V$, $\xi_{A_1}$ and $\xi_{A_3}$ were modeled by the single form
factor, as indicated in eq.~(\ref{iwr}),
\begin{equation}
{\cal F}(\eta)={\cal F}(1)[1-a^2(\eta-1)+b(\eta-1)^2]
\label{mi}
\end{equation}
with the parameters $a^2=0.84$, $b=0$ for a linear fit and $a^2=0.92$,
$b=0.15$ for a quadratic fit to experimental data. The normalization
${\cal F}(1)=\eta_A\xi(1)+{\cal O}((\Lambda/M)^2)$, where $\eta_A$ is
a perturbatively calculable quantity, takes the value ${\cal F}(1)=0.93$
\cite{LNN}, ${\cal F}(1)=0.89$ \cite{SUV} or ${\cal F}(1)=0.96$ \cite{M}.
In Ref.~\cite{AGS} the single form factor was expressed as the overlap
integral of the $B$ and $D^{(*)}$ meson wave functions derived from the
Bethe-Salpeter equation, and was parametrized by a similar formula to
eq.~(\ref{mi}),
\begin{equation}
{\cal F}(\eta)=\eta_A\left[1-\frac{\rho_1^2}{\eta_A}(\eta-1)
+c(\eta-1)^{3/2}\right]\;,
\label{28}
\end{equation}
for the constants $\eta_A=0.9942$, $\rho_1^2=1.279$ and $c=0.91$. The
method of overlap integrals leads to the expression \cite{NR}
\begin{equation}
{\cal F}(\eta)=\frac{2}{\eta+1}\exp\left[-2(\rho_2^2-1)\frac{\eta-1}
{\eta+1}\right]\;,
\label{29}
\end{equation}
for the constant $\rho_2=1.19$. In Ref.~\cite{GGL} the best fit to the
CLEO data \cite{B} gives the form factor
\begin{equation}
{\cal F}(\eta)=1-0.81(\eta-1)\;.
\label{30}
\end{equation}
Comparing to our results, we find that eqs.~(\ref{mi})-(\ref{30})
are in fact close to or larger than $\xi_+$ as shown in fig.~6.
Since our form factors involved in the decay $B\to D^*\ell\nu$ are smaller,
the prediction of $|V_{cb}|$ is of course
larger. If substituting $\xi_+$ for $\xi_V$, $\xi_{A_1}$ and $\xi_{A_3}$ in
eq.~(\ref{dd}), we shall have $|V_{cb}|\approx 0.043\times 0.9=0.039$
extracted from the data, which then locates in the above range.

We explain why our form factors are smaller than those in the literature. The
reason is attributed to the choice of the decay constants $f_B=0.12$ GeV
and $f_D=0.14$ GeV at the beginning of the analysis. If $f_B$ and $f_D$
increase to 0.13 and 0.15 GeV, respectively, $f_{D^*}$ will become 0.138 GeV
because of the ratio $f_{D^*}/f_D=0.92$. Here we suppose that the shape
parameters $C_B$, $C_D$ and $C_{D^*}$ change only slightly. Then $|V_{cb}|$
decreases to 0.037 in order to maintain the height of the
spectrum. We have confirmed this argument by following Steps 1 to 4
explicitly as stated above. Therefore, the best conclusion for our
study is that we have disentangled the task of determining $|V_{cb}|$ to
the extent that $|V_{cb}|$ is given, in terms of $f_B$ and $f_D$, by
\begin{equation}
|V_{cb}|=0.043\times \left(\frac{0.12\;\;{\rm GeV}}{f_B}\right)\times
\left(\frac{0.14\;\;{\rm GeV}}{f_D}\right)\;,
\end{equation}
for $f_B$ and $f_D$ varying around 0.12 and 0.14 GeV, respectively.
Once the precise measurement of the decay constants $f_B$ and $f_D$ is
performed, the CKM matrix element $|V_{cb}|$ can be fixed uniquely.

At last, we compute the branching ratio ${\cal B}({\bar B}^0\to D^+D_s^-)$
inserting the parameters in eq.~(\ref{fp}), and compare results with the
data $8.0\times 10^{-3}$ \cite{PDG}, if ignoring the errors. The expression
for the decay rate is written as
\begin{eqnarray}
\Gamma({\bar B}^0\to D^+D_s^-)&=&\frac{1}{32\pi}G_F^2|V_{cb}|^2|V_{cs}|^2
f_{D_s}^2M_B^3(1-r^2)^2\sqrt{\eta_{\rm max}^2-1}
\nonumber \\
& &\times\left[\frac{(1+r)^2-r'^2}{1+r}\xi_+
(\eta_{\rm max})-\frac{(1-r)^2-r'^2}{1-r}
\xi_-(\eta_{\rm max})\right]^2\;,
\nonumber \\
& &
\end{eqnarray}
for the CKM matrix element $|V_{cs}|=1$ \cite{PDG}, $r'=M_{D_s}/M_B$ and
$\eta_{\rm max}=(1+r^2-r'^2)/(2r)$. We obtain the
$D_s$ meson decay constant $f_{D_s}=0.137$ GeV, or in terms of the ratio,
$f_{D_s}/f_D=0.98$. Because of the large errors associated with the data,
we do not compare this ratio with those from the lattice calculation
\cite{BLS} and from QCD sum rules \cite{D}, which are about 1.1 and 1.2,
respectively. However, our prediction is in agreement with the simple formula
\cite{O}
\begin{equation}
\frac{f_{D_s}}{f_D}=\left(\frac{M_D}{M_{D_s}}\right)^2\frac{m_c+m_s}{
m_c+m_d}\approx 0.98
\label{ar}
\end{equation}
for the current quark masses $m_d=10$ MeV, $m_s=150$ MeV and
$m_c=1.5$ GeV. Equation (\ref{ar}) was obtained using general arguments
from the Wigner-Eckart theorem and assuming that chiral symmetry
is only broken by quark mass terms.
\vskip 2.0cm

\section{Nonfactorizable Contributions}
\vskip 0.5cm

In the study of $B\to D^{(*)}$ decays, we have considered only
factorizable contributions from external $W$-emission diagrams as in fig.~1a.
For the nonleptonic decay ${\bar B}^0\to D^+\pi^-$, there are also
nonfactorizable contributions from $W$-exchange diagrams as shown in fig.~7a.
To justify our study, we should have a convincing argument
that such nonfactorizable contributions are indeed unimportant. A simple
investigation shows that the PQCD formalism can be applied to fig.~7a
equally well, with the following modifications:

\noindent
1. The color flow in fig.~7a differs from that in fig.~1a. This distinction
leads to a factor $1/3$ for nonfactorizable contributions.

\noindent
2. From the viewpoint of factorization theorems, fig.~7a is a crossing in
the $s$ and $t$ channels of fig.~1a, excluding the color flow. That
is, these two diagrams are similar to each other, except the interchange of
the $B$ meson and pion kinematic variables. Hence, the hard gluon propagator
in fig.~7a is proportional to $1/x_3x_2$, $x_3$ being the momentum fraction
of the pion, which comes from the replacement of $x_1$ by $x_3$ in the gluon
propagator $1/x_1x_2$ associated with fig.~1a. Since $x_1$ and $x_3$ are of
order 0.05 and 0.5, respectively, from the $B$ meson and pion wave functions,
the interchange leads to a factor $1/10$.

\noindent
Certainly, the presence of $k_T^2$ in the hard scatterings moderates
the difference. Therefore, we estimate that nonfactorizable contributions
from $W$-exchange diagrams are roughly 5\% of factorizable ones, and the
neglect of fig.~7a is reasonable.

With the parameters determined in Sect. 3, we can also study charged $B$
meson decays, such as $B^-\to D^0\pi^-$, based on the PQCD formalism
employed here. In this case there are additional contributions from
internal $W$-emission diagrams as shown in fig.~7b. Following the similar
reasoning, the hard scattering associated with fig.~7b, which is proportional
to $1/x_1x_3$, is obtained by interchanging the kinematical variables of the
$D$ meson and of the pion. This interchange gives a factor $0.2/0.5=2/5$.
We estimate that contributions from internal $W$-emission diagrams,
combined with the color-suppressing factor 1/3, are roughly 15\% of
factorizable ones, which are of course sizable. This conclusion is
consistent with predictions from the BSW method \cite{A}. We shall discuss
these subjects in details in a seperate work \cite{WYL}.

\vskip 2.0cm
\section{Conclusion}
\vskip 0.5cm

In this paper we have fixed the $D^{(*)}$ meson wave function using the
experimental inputs from the nonleptonic decays $B\to D^{(*)}\pi$,
and evaluated the spectrum of the semileptonic decay $B\to D^*\ell\nu$
at large velocity transfer, from which the CKM matrix element
$|V_{cb}|=0.043$ is extracted. The form factors involved in $B\to D^{(*)}$
transitions are obtained, and compared to the predictions from HQS combined
with ${\cal O}(\alpha_s)$ and ${\cal O}(1/M)$ corrections \cite{N}, from
the overlap integrals of heavy meson wave functions \cite{NR,AGS}, and
from the data fitting \cite{B}. The value of $|V_{cb}|$ extracted here is
larger than those in the literature \cite{AGS,B,GGL}, and the reason is
that we have adopted the smaller decay constants $f_B$ and
$f_D$. A precise measurement of $f_B$ and $f_D$ in the future will remove
this ambiguity.

We emphasize that the behaviors of all the transition form factors are
derived from the single $D^{(*)}$ meson wave function that is fixed at
specific kinematic points, without resorting to a model for each form factor,
such as the algebraic forms employed in \cite{LNN}, the pole forms in
\cite{BSW,KS} and the exponential forms in \cite{ISGW}. Note that all the
above model form factors are larger than $\xi_+$ presented in this work.
Hence, they lead to smaller values of $|V_{cb}|$, ranging from 0.032 to
0.038. Our formalism, based on the parameters in eq.~(\ref{fp}), can be used
to study $B$ meson decays, especially the nonleptonic cases with
nonfactorizable contributions, in a less ambiguous way \cite{WYL}.
\vskip 0.5cm

We thank H.Y. Cheng and P. Kroll for useful discussions.
This work was supported by National Science council of R.O.C. under
the Grant No. NSC-85-2112-M-194-009.
\vskip 2.0cm

\centerline{\large \bf Appendix}
\vskip 0.3cm

In this appendix we present the explicit expression of the exponent $s(k,b)$
appearing in eq.~(\ref{wp}). The full expression, instead of the first
six terms \cite{LS}, is adopted in this paper. It is given, in terms of
the variables,
\begin{eqnarray}
{\hat q} \equiv  {\rm ln}\left(k/\Lambda\right),\;\;\;\;\;
{\hat b} \equiv {\rm ln}(1/b\Lambda)
\end{eqnarray}
by \cite{L1}
\begin{eqnarray}
s(k,b)&=&\frac{A^{(1)}}{2\beta_{1}}\hat{q}\ln\left(\frac{\hat{q}}
{\hat{b}}\right)-
\frac{A^{(1)}}{2\beta_{1}}\left(\hat{q}-\hat{b}\right)+
\frac{A^{(2)}}{4\beta_{1}^{2}}\left(\frac{\hat{q}}{\hat{b}}-1\right)
\nonumber \\
& &-\left[\frac{A^{(2)}}{4\beta_{1}^{2}}-\frac{A^{(1)}}{4\beta_{1}}
\ln\left(\frac{e^{2\gamma-1}}{2}\right)\right]
\ln\left(\frac{\hat{q}}{\hat{b}}\right)
\nonumber \\
& &+\frac{A^{(1)}\beta_{2}}{4\beta_{1}^{3}}\hat{q}\left[
\frac{\ln(2\hat{q})+1}{\hat{q}}-\frac{\ln(2\hat{b})+1}{\hat{b}}\right]
\nonumber \\
& &+\frac{A^{(1)}\beta_{2}}{8\beta_{1}^{3}}\left[
\ln^{2}(2\hat{q})-\ln^{2}(2\hat{b})\right]
\nonumber \\
& &+\frac{A^{(1)}\beta_{2}}{8\beta_{1}^{3}}
\ln\left(\frac{e^{2\gamma-1}}{2}\right)\left[
\frac{\ln(2\hat{q})+1}{\hat{q}}-\frac{\ln(2\hat{b})+1}{\hat{b}}\right]
\nonumber \\
& &-\frac{A^{(1)}\beta_{2}}{16\beta_{1}^{4}}\left[
\frac{2\ln(2\hat{q})+3}{\hat{q}}-\frac{2\ln(2\hat{b})+3}{\hat{b}}\right]
\nonumber \\
& &-\frac{A^{(1)}\beta_{2}}{16\beta_{1}^{4}}
\frac{\hat{q}-\hat{b}}{\hat{b}^2}\left[2\ln(2\hat{b})+1\right]
\nonumber \\
& &+\frac{A^{(2)}\beta_{2}^2}{1728\beta_{1}^{6}}\left[
\frac{18\ln^2(2\hat{q})+30\ln(2\hat{q})+19}{\hat{q}^2}
-\frac{18\ln^2(2\hat{b})+30\ln(2\hat{b})+19}{\hat{b}^2}\right]
\nonumber \\
& &+\frac{A^{(2)}\beta_{2}^2}{432\beta_{1}^{6}}
\frac{\hat{q}-\hat{b}}{\hat{b}^3}
\left[9\ln^2(2\hat{b})+6\ln(2\hat{b})+2\right]\;.
\label{sss}
\end{eqnarray}
The above coefficients $\beta_{i}$ and $A^{(i)}$ are
\begin{eqnarray}
& &\beta_{1}=\frac{33-2n_{f}}{12}\;,\;\;\;\beta_{2}=\frac{153-19n_{f}}{24}\; ,
\nonumber \\
& &A^{(1)}=\frac{4}{3}\;,
\;\;\; A^{(2)}=\frac{67}{9}-\frac{\pi^{2}}{3}-\frac{10}{27}n_
{f}+\frac{8}{3}\beta_{1}\ln\left(\frac{e^{\gamma_E}}{2}\right)\; ,
\end{eqnarray}
where $\gamma_E$ is the Euler constant.

Note that $s$ is defined for ${\hat q}\ge {\hat b}$, and set to zero for
${\hat q}<{\hat b}$. As a similar treatment, the complete Sudakov factor
$\exp(-S)$ is set to unity, if $\exp(-S)>1$, in the numerical analysis.
This corresponds to a truncation at large $k_T$, which spoils the
on-shell requirement for the light valence quarks. The quark lines with
large $k_T$ should be absorbed into the hard scattering amplitude, instead
of the wave functions. An explicit examination  shows that the partial
expression including only the first six terms gives predictions smaller than
those from the full expression by less than 5\% \cite{L1}.

\newpage
\cl{\large \bf Figure Captions}
\vskip 0.5cm

\noindent
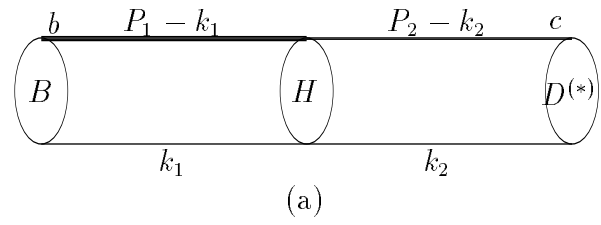
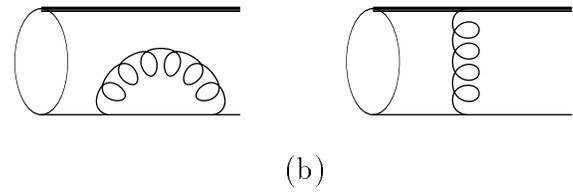
{\bf Fig. 1.} (a) Factorization of $B\to D^{(*)}$ transitions. (b)
${\cal O}(\alpha_s)$ corrections to wave functions.
\vskip 0.5cm

\noindent
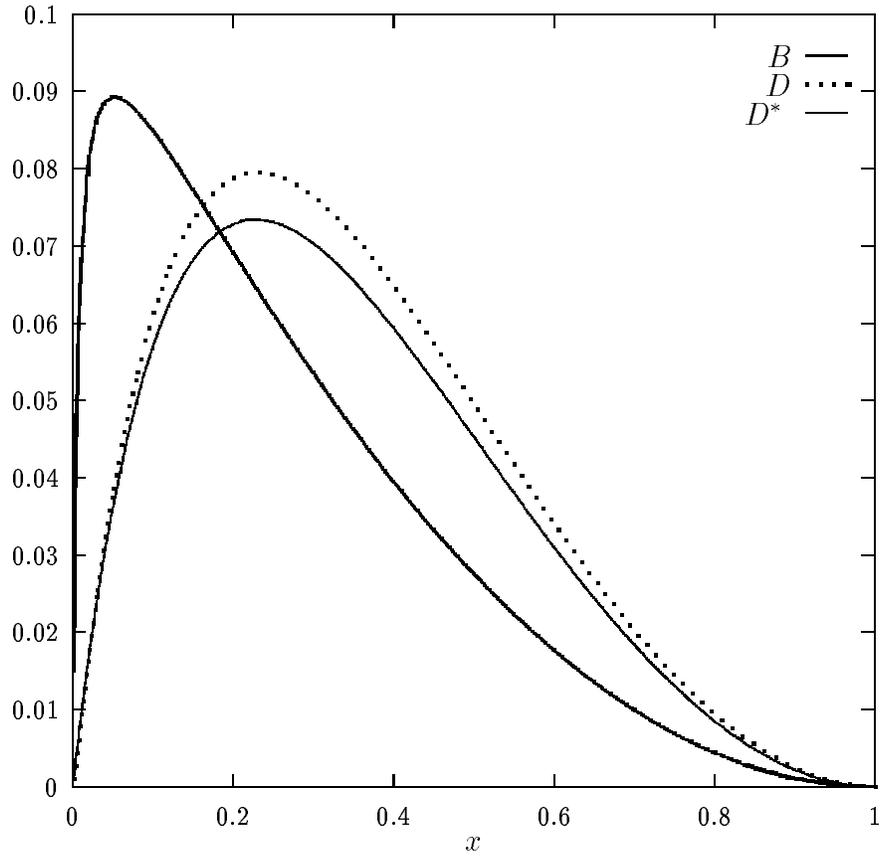
{\bf Fig. 2.} Dependence of the $B$ and $D^{(*)}$ meson wave functions
on the momentum fraction $x$.

\noindent
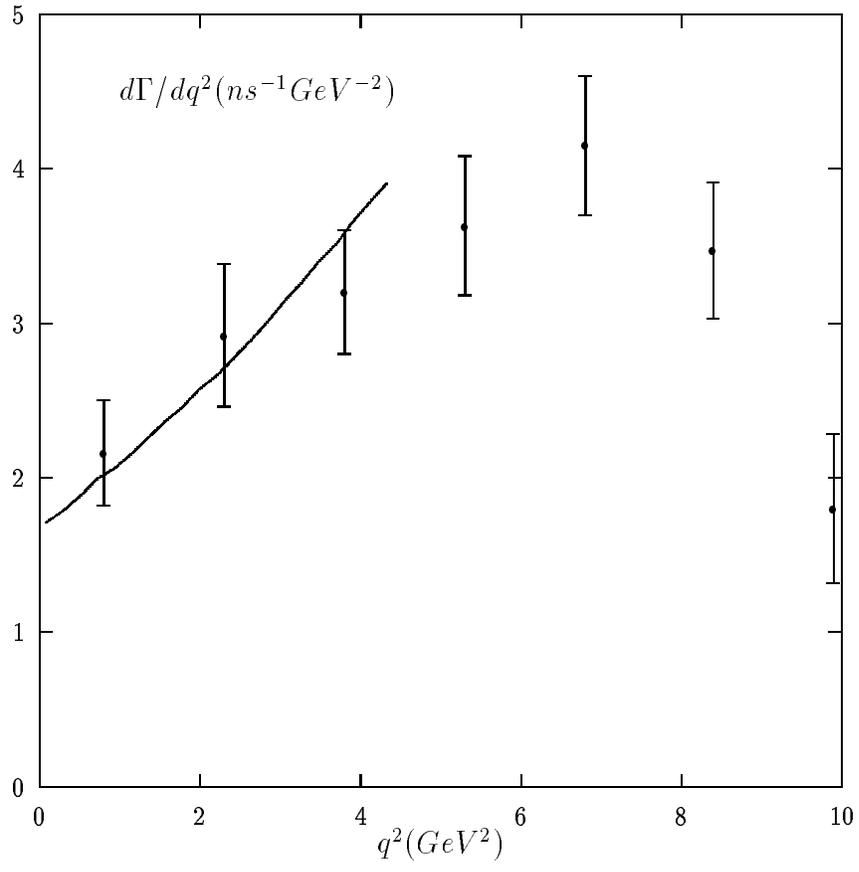
{\bf Fig. 3.} The spectrum $d\Gamma/dq^2$ of the semileptonic decay ${\bar
B}^0\to D^{*+}\ell^-{\bar \nu}$.
\vskip 0.5cm

\noindent
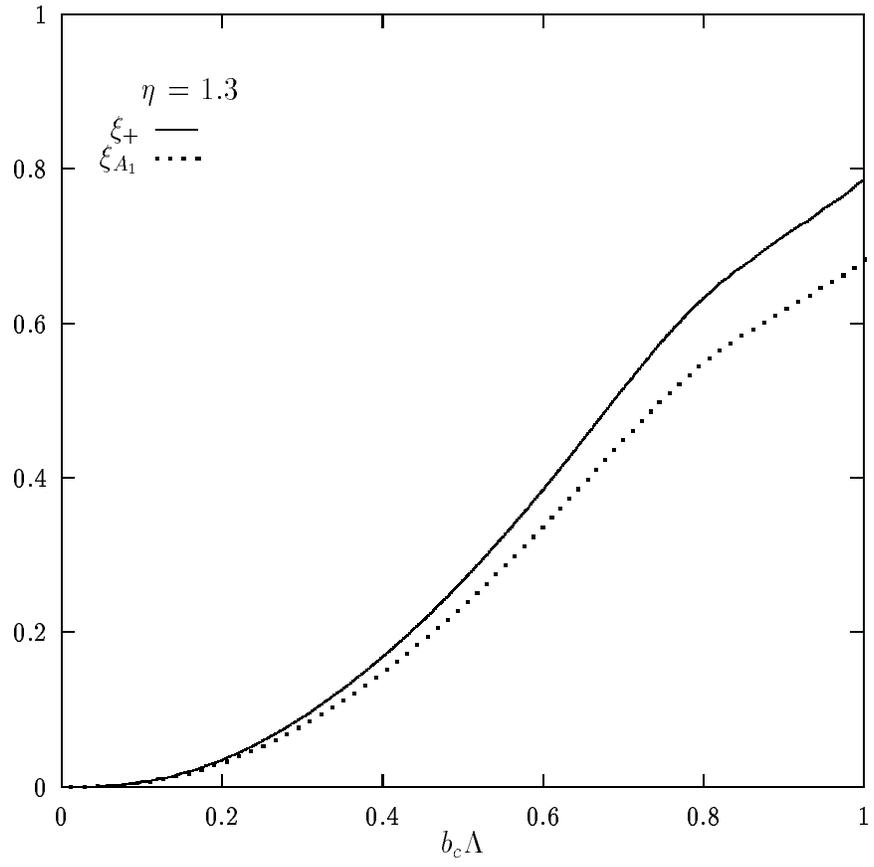
{\bf Fig. 4.} Dependence of $\xi_+$ and $\xi_{A_1}$
on the cutoff $b_c$  for $\eta=1.3$.
\vskip 0.5cm

\noindent
{\bf Fig. 5.} Dependence of (a) $\xi_+$, $\xi_V$, $\xi_{A_1}$ and $\xi_{A_3}$
and of (b) $\xi_-$ and $\xi_{A_2}$ on $\eta$.
\vskip 0.5cm

\noindent
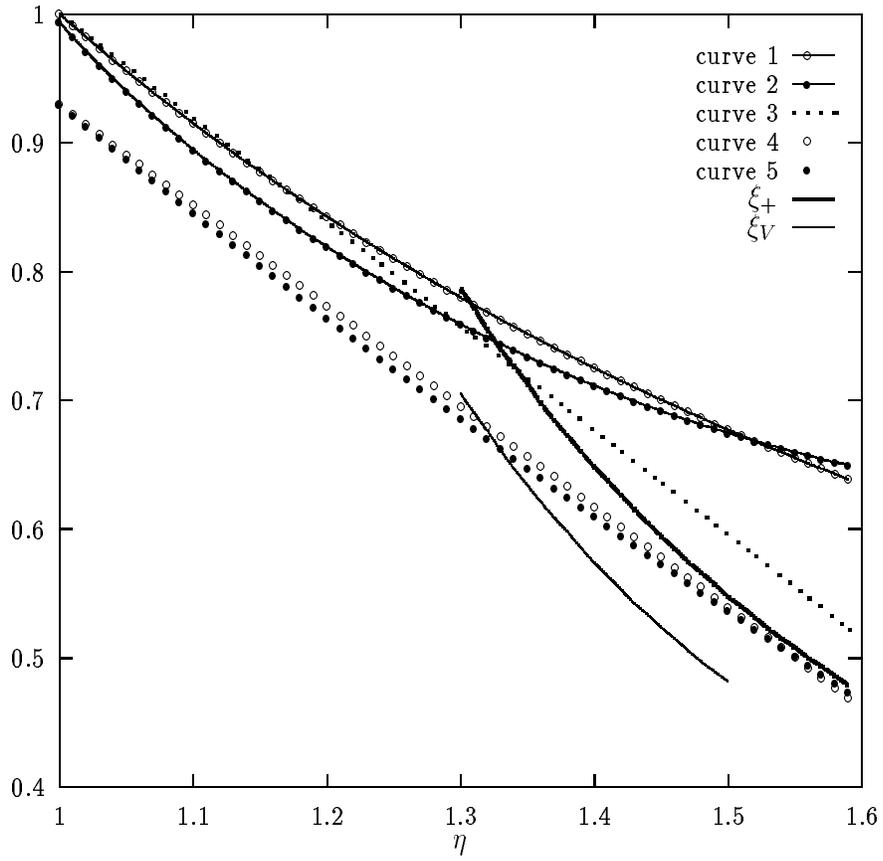
{\bf Fig. 6.} Dependence on $\eta$ of (1) eq.~(\ref{29}), (2) eq.~(\ref{28}),
(3) eq.~(\ref{30}), (4) eq.~(\ref{mi}) for the linear fit with
${\cal F}(1)=0.93$, and (5) eq.~(\ref{mi}) for the quadratic fit with
${\cal F}(1)=0.93$. The curves of $\xi_+$ and $\xi_V$
are also shown.
\vskip 0.5cm

\noindent
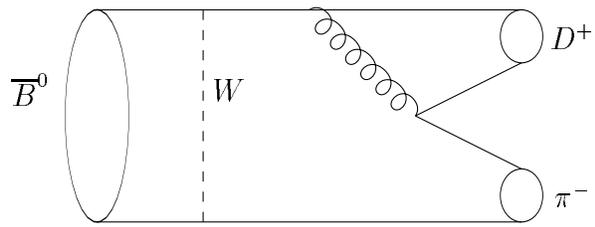
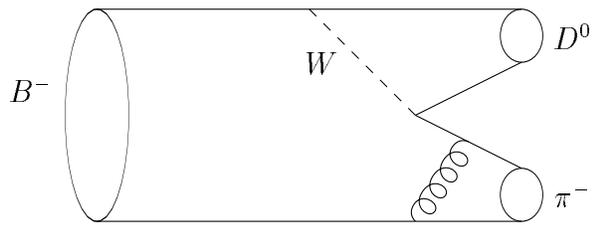
{\bf Fig. 7.} (a) $W$-exchange and (b) internal $W$-emission ${\cal O}
(\alpha_s)$ diagrams.


\newpage


\newpage
\begin{thebibliography}{99}
\bibitem{LY1} H.-n. Li and H.L. Yu, Phys. Rev. Lett. 74, 4388 (1995);
Phys. Lett. B (1995); H.-n. Li, Phys. Lett. B348, 597 (1995).
\bibitem{L1} H.-n. Li, Phys. Rev. D52, 3958 (1995).
\bibitem{L2} H.-n. Li, preprint CCUTH-95-02.
\bibitem{LY2} H.-n. Li and H.L. Yu, preprint CCUTH-95-04.
\bibitem{S} F. Schlumpf, talk given at XIVth European Conference on
Few-Body problems in Physics, Amsterdam, The Netherlands, 1993 (unpublished).
\bibitem{A} M.S. Alam {\it el al.} (CLEO Collaboration), Phys. Rev. D50,
43 (1994).
\bibitem{IW} N. Isgur and M.B. Wise, Phys. Lett. B232, 113 (1989);
237, 527 (1990).
\bibitem{N} M. Neubert, Phys. Lett. B264, 455 (1991).
\bibitem{NR} M. Neubert and V. Rieckert, Nucl. Phys. B382, 97 (1992).
\bibitem{AGS} A. Abd El-Hady, K.S. Gupta, A.J. Sommerer, J. Spence and
J.P. Vary, Phys. Rev. D51, 5245 (1995).
\bibitem{B} B. Barish {\it et al.} (CLEO Collaboration), Phys. Rev.
D51, 1014 (1995).
\bibitem{PDG} Review of Particle Properties, Phys. Rev. D50 (1994).
\bibitem{BLS} C. Bernard, talk presented at the CTEQ summer school on QCD
analysis and phenomenology, Missouri, 1994 (unpublished);
C. Bernard, J. Labrenz and A. Soni, Phys. Rev. D49, 2536 (1994).
\bibitem{BSS} D. Bortoletto and S. Stone, Phys. Rev. Lett. 65, 2951 (1990).
\bibitem{ST} K. Berkelman and S. Stone, Annu. Rev. Nucl. Part. Sci. 41,
1 (1991).
\bibitem{BSW} M. Bauer, B. Stech and M. Wirbel, Z. Phys. C29, 637 (1985);
34, 103 (1987).
\bibitem{AN} A. Abada {\it et al.}, Nucl. Phys. B376, 172 (1992);
M. Neubert, Phys. Rev. D46, 1076 (1992).
\bibitem{GGL} C.G. Boyd, B. Grinstein and R.F. Lebed, Phys. Lett. B353, 306
(1995).
\bibitem{LNN} M. Neubert, Phys. Rep. 245, 261 (1994); Z. Ligeti, Y. Nir
and M. Neubert, Phys. Rev. D49, 1302 (1994).
\bibitem{SUV} M. Shifman, N. Uraltsev and A. Vainshtein, University of
Minnesota Report No. TPI-MINN-94/13-T (unpublished).
\bibitem{M} T. Mannel, Phys. Rev. D50, 428 (1994).
\bibitem{D} C.A. Dominguez, in Proceedings of the Third Workshop on
the Tau-Charm Factory, Spain (1993).
\bibitem{O} R.J. Oakes, Phys. Rev. Lett. 73, 381 (1994).
\bibitem{WYL} C.Y. Wu, T.W. Yeh and H.-n. Li, in preparation.
\bibitem{KS} J.G. K\"orner and G.A. Schuler, Z. Phys. C38, 511 (1989).
\bibitem{ISGW} N. Isgur, D, Scora, B. Grinstein and M.B. Wise, Phys. Rev.
D39, 799 (1989).
\bibitem{LS} H.-n. Li and G. Sterman, Nucl. Phys. B381, 129 (1992);
H.-n. Li, Phys. Rev. D48, 4243 (1993).

\end{thebibliography}
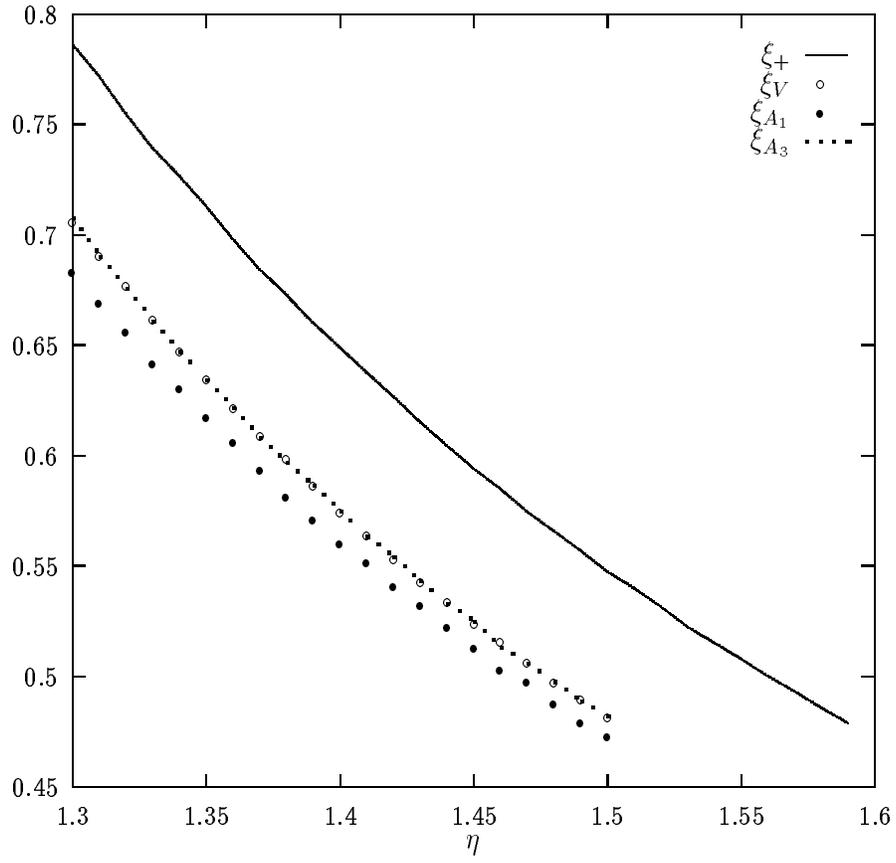
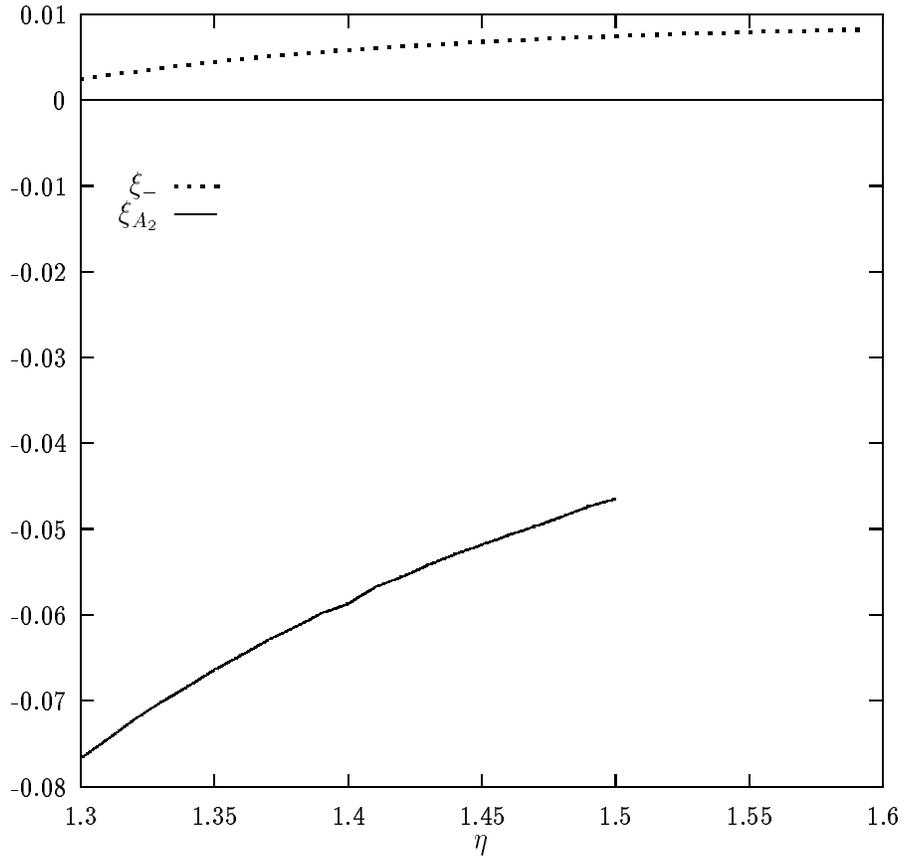
\end{document}